\documentclass[runningheads]{llncs}
\usepackage[T1]{fontenc}
\usepackage{graphicx}
\usepackage{tabularx}
\usepackage{booktabs}
\usepackage{subfigure}
\usepackage{multirow}
\usepackage{enumitem}
\usepackage{wrapfig}
\usepackage{marvosym}
\usepackage{xcolor}
\begin{document}
\title{How to Forget Clients in Federated Online Learning to Rank?}

\author{Shuyi Wang\inst{1}\textsuperscript{(\Letter)}\orcidID{0000-0002-4467-5574} \and
Bing Liu\inst{2}\orcidID{0000-0002-7858-7468} \and
Guido Zuccon\inst{1}\orcidID{0000-0003-0271-5563}}
\authorrunning{S. Wang et al.}
\institute{The University of Queensland, Australia \\ \email{\{shuyi.wang, g.zuccon\}@uq.edu.au} \and CSIRO, Australia \\ \email{liubing.csai@gmail.com}}
%%

%\institute{Princeton University, Princeton NJ 08544, USA \and
%	Springer Heidelberg, Tiergartenstr. 17, 69121 Heidelberg, Germany
%	\email{lncs@springer.com}\\
%	\url{http://www.springer.com/gp/computer-science/lncs} \and
%	ABC Institute, Rupert-Karls-University Heidelberg, Heidelberg, Germany\\
%	\email{\{abc,lncs\}@uni-heidelberg.de}}

\maketitle              % typeset the header of the contribution
\begin{abstract}
Data protection legislation like the European Union's General Data Protection Regulation (GDPR) establishes the \textit{right to be forgotten}: a user (client) can request contributions made using their data to be removed from learned models. In this paper, we study how to remove the contributions made by a client participating in a Federated Online Learning to Rank (FOLTR) system. In a FOLTR system, a ranker is learned by aggregating local updates to the global ranking model. Local updates are learned in an online manner at a client-level using queries and implicit interactions that have occurred within that specific client. By doing so, each client's local data is not shared with other clients or with a centralised search service, while at the same time clients can benefit from an effective global ranking model learned from contributions of each client in the federation.

In this paper, we study an effective and efficient unlearning method that can remove a client's contribution without compromising the overall ranker effectiveness and without needing to retrain the global ranker from scratch. A key challenge is how to measure whether the model has unlearned the contributions from the client $c^*$ that has requested removal.
For this, we instruct $c^*$ to perform a poisoning attack (add noise to this client updates) and then we measure whether the impact of the attack is lessened when the unlearning process has taken place. Through experiments on four datasets, we demonstrate the effectiveness and efficiency of the unlearning strategy under different combinations of parameter settings.

\end{abstract}
\section{Introduction}

In Online Learning to Rank (OLTR), ranking models keep evolving by being updated using users' implicit feedback (e.g. click data) on the relevance between queries and documents in an online manner~\cite{yue2009interactively,oosterhuis2018differentiable,wang2019variance,zhuang2020counterfactual,ai2021unbiased}.
Though OLTR provides a mechanism to learn effective ranking models, it also raises privacy concerns as it requires to collect users' interaction data to the server for centralised training.
This paradigm is thus not suitable to privacy-sensitive situations where users do not want to share their data.
To support privacy protection, a new OLTR setting which characterizes no data sharing -- Federated Online Learning to Rank (FOLTR) -- has been explored~\cite{kharitonov2019federated,wang2021federated}. 
In this setting, each client (i.e. user) exploits its own data to update the ranker locally; it then sends the ranker update, instead of its data, to a central server. The server aggregates the received updates to derive an updated global ranker, which is subsequently broadcast to each client. This federated paradigm is suitable to both web-scale training of rankers, where many clients are involved (cross-device FOLTR\footnote{\label{fl}Similar concepts as cross-device and cross-silo Federated Learning~\cite{kairouz2021advances}}), and the institutional training of rankers, where only a few institutions or organisations are involved (cross-silo FOLTR\footref{fl}).

A considerate federated learning system should consider the possibility for clients to leave the federation and request the contributions of their data erased~\cite{liu2021federaser,DBLP:journals/corr/abs-2003-10933,DBLP:conf/infocom/LiuXYWL22}. 
This possibility -- dubbed the \textit{right to be forgotten} -- is contemplated in modern data protection legislation, such as in the General Data Protection Regulation (GDPR) emanated by the European Union~\cite{de2020european}. However, the design of existing FOLTR systems is defective as lacking an \textit{unlearning} mechanism to forget certain users' contributions.
An effective and efficient unlearning mechanism is not straightforward to design.
A naive way is to ask all remaining clients to re-execute the training of ranker from scratch. This carries implications in terms of disruption of service and comes with large computational costs, even if the update was done in an offline manner (counterfactually on log data stored in each client), rather than in an online manner (which in turns is impractical as it needs the users to interact again with the search results). 
Therefore, a more reasonable unlearning mechanism for FOLTR is necessary, but has not been studied. In this paper, we aim to fill this gap and provide an initial investigation of unlearning mechanisms for FOLTR.

In particular, we strive to overcome two main challenges. 
The first challenge is how to \textit{efficiently} unlearn without requiring an unreasonable amount of additional computation. Also, the obtained new ranker is expected to have comparable effectiveness to the one retrained from scratch. 
The second challenge is how to evaluate the \textit{effectiveness} of an unlearning method.
In a FOLTR system with many clients, the effect on ranker's effectiveness of removing a client can be marginal, or even unnoticeable. A natural question from a user leaving the federation is: how can it be proven that the impact of my data on the ranker has been erased? An adequate evaluation method is then required to verify whether an unlearning process is effective in forgetting.

To address these challenges and facilitate future studies, we build the first benchmark for unlearning in FOLTR.
We adapt an effective unlearning technique emerging from the general federated machine unlearning field (FedEraser~\cite{liu2021federaser}) to the context of FOLTR with adaptation into online training. 
In our method, some historical updates are stored in the local devices and re-used to help retrain a new ranker with much less additional computation cost.
In addition, for evaluation, we adopt a poisoning attack method~\cite{wang2023analysis} to magnify and control the effect of the client leaving the federation.
Through extensive empirical experimentation across four learning-to-rank datasets, we study the effectiveness and efficiency of the unlearning method and the factors influencing its performance. 
The utility of our evaluation method is also verified.
This paper is the first study that investigates unlearning in FOLTR systems, where it is not clear that advances in general federated learning translate to similar improvements\footnote{For example, many methods that are successful for dealing with non identical and independently distributed data (non-IID) in general federated learning do not work in FOLTR~\cite{wang2022non}.}. 
This is because of the significant difference between FOLTR from general classification tasks, e.g. ranking vs. classification, online learning vs. offline learning\footnote{In federated learning, the local model can be trained on the local data repeatedly across several epochs, while in FOLTR, training data is acquired in real time as user interactions occur and it cannot be repeated and reused (e.g., a user cannot be asked to submit the same query they did in the past, and perform the same interactions). }, implicit user feedback vs. ground-truth labels, etc.
In addition to the impact on FOLTR, our benchmark can enrich the task-level diversity for the evaluation of general federated unlearning methods.

\section{Related Work}

\textbf{Federated Online Learning to Rank.}
FOLTR systems consider a decentralized OLTR~\cite{hofmann2013reusing,schuth2016multileave,zhao2016constructing,oosterhuis2018differentiable,wang2018efficient,zhuang2020counterfactual,zhuang2022reinforcement,jia2022learning} scenario where data owners (clients) collaboratively train a ranker in an online manner under the coordination of a central server without the need of sharing their data. 
Though FOLTR is still largely unexplored, few existing works have established an initial landscape of this research area.
The Federated OLTR with Evolutionary Strategies (FOLtR-ES)~\cite{kharitonov2019federated} method was the first FOLTR system proposed in the literature. FOLtR-ES extends the centralised OLTR scenario into federated setting~\cite{mcmahan2017communication} and uses Evolution Strategies as optimization method~\cite{salimans2017evolution}. 
While FOLtR-ES performs well on small-scale datasets under specific evaluation metrics, its effectiveness does not generalize to large-scale datasets and standard OLTR metrics~\cite{wang2021federated}.
To improve the effectiveness of FOLTR, Wang et al.~\cite{wang2021effective} proposed an alternative method named FPDGD~\cite{wang2021effective}, which leverages the state-of-the-art OLTR method, the Pairwise Differentiable Gradient Descent (PDGD)~\cite{oosterhuis2018differentiable}, and adapts it to the Federated Averaging (FedAvg) framework~\cite{mcmahan2017communication}. 
FPDGD's effectiveness is comparable to that of centralized OLTR methods and is currently the state-of-the-art FOLTR method.
Though, we noticed that Wang et al.~\cite{wang2022non} pointed out that FPDGD's effectiveness has been shown to deteriorate if data is not distributed identically and independently (non-IID) across clients.

This paper extends the landscape of existing works by adding an unlearning mechanism to a FOLTR system.
In our experiments we use FPDGD~\cite{wang2021effective} as the base FOLTR system. 
To avoid entangling different sources of challenges, we do not consider non-IID settings~\cite{wang2022non} and  privacy-preserving mechanisms~\cite{wang2021effective} within unlearning in our experiments.
The study of unlearning across these more complex experimental settings is left for future work.

\textbf{Machine Unlearning.} Our work is related to machine unlearning~\cite{bourtoule2021machine,nguyen2022survey}, which pertains to the removal of any evidence of a chosen data point from the model, a process commonly known as selective amnesia. Except ensuring the removal of certain data points from the model being the primary objective, the unlearning procedure should also do not affect the model's effectiveness. Machine unlearning has been explored in both centralised~\cite{DBLP:conf/sp/CaoY15,bourtoule2021machine} and federated settings~\cite{liu2021federaser,halimi2022federated,wang2022federated,che2023fast}, but never for FOLTR -- this is a novel contribution of our work. 

Methods in machine unlearning can be broadly classified into two families: exact unlearning and approximate unlearning. Exact unlearning methods are designed to provide a theoretical guarantee that the methods can completely remove the influence of the data to be forgotten~\cite{DBLP:journals/ml/BaumhauerSZ22,DBLP:conf/icml/BrophyL21,DBLP:journals/cluster/ChenXXZ19,DBLP:conf/nips/GinartGVZ19,mahadevan2021certifiable}; but a limitation of these methods is that they can only be applied to simple machine learning models.
Approximate unlearning methods, on the other hand,  are characterized by higher efficiency, which is achieved by relying on specific assumptions regarding the accessibility of training information, and by permitting a certain amount of reduction in the model's effectiveness~\cite{DBLP:journals/access/AldaghriMB21,DBLP:conf/ccs/Chen000H022,chundawat2023zero}.
These methods can be used on more complex machine learning models (e.g., deep neural networks). 
The existing unlearning methods for federated learning belong to the approximate unlearning family~\cite{liu2021federaser,DBLP:journals/corr/abs-2003-10933,DBLP:conf/infocom/LiuXYWL22}: our method builds upon one such methods, FedEraser~\cite{liu2021federaser}.
However, most of these methods are applied to classification tasks and no previous work considers either ranking or FOLTR systems.

\textbf{Evaluation of Machine Unlearning.}
A key challenge posed by the task of unlearning in the federated learning context is how to evaluate whether an unlearning method has successfully removed the contributions from the client that requested to leave the federation. This evaluation is at times termed as unlearning verification~\cite{nguyen2022survey}, which specifically aims to certify that the unlearned model is unrelated to the data that needed to be removed. 
Due to the stochastic nature of the training for many machine learning models, it is difficult to distinguish the individual clients' models and their unlearned counterparts after a certain group of data is removed. 

To evaluate the effectiveness of unlearning, a group of methods leverage membership inference attacks~\cite{liu2021federaser,chen2021machine,hu2022membership}, i.e. a kind of attack method that predicts whether a data item belongs to the training data of a certain model~\cite{shokri2017membership}: it is then straightforward to adapt this type of attacks to verify if the unlearned samples participated in the unlearning process. However, conducting successful membership inference attacks needs subtle design and training of the inference model. The effectiveness of such attacks on FOLTR is unknown as it is an unexplored area. Thus, we do not consider this type of verification in our study.

Another type of methods is inspired from the idea of poisoning attacks; in our experiments, we embrace this direction for evaluating unlearning. 
These methods add arbitrary noise~\cite{yuan2023federated} or backdoor triggers~\cite{gao2022verifi,sommer2022athena,wu2022federated,halimi2022federated} to the data that is needed to be deleted, with the aim of manipulating the effectiveness of the trained model. After the unlearning of poisoned datasets has taken place, the impact from the arbitrary poisoning or backdoor triggers should be reduced in the unlearned model: the extent of this reduction (and thus model gains) determines the extent of the success of unlearning.

A significant difference between our work and existing unlearning works is that we consider the context of FOLTR, which has specific challenges not present in common classification tasks (e.g., ranking vs. classification, online learning vs. offline learning, etc.). Thus, methods proposed in the general machine learning community may not work in FOLTR. 
Instead of diving into a specific problem, we establish the first unlearning benchmark -- including both unlearning method and evaluation setting -- for FOLTR. By doing so, we facilitate future works which focus on the specific challenges of unlearning in FOLTR.

\section{Methodology}
\label{method}

\begin{table}[t]
	\centering
	\caption{Notation used in this paper. 	\label{tbl_notation} }
	\resizebox{0.9\columnwidth}{!}{		
		\begin{tabular}{ll}
			\toprule
			\bf Symbol  & \bf Description \\
			\midrule
			$c_i$        & client $i$ in the FOLTR system           \\
			$c^*$        & the unlearned client that requested removal from the FOLTR system           \\
			$n_i$        & number of local updates for client $i$, before unlearning takes place\\ 
			$n'_i$        & number of local updates for client $i$, during the unlearning process \\ 
			$T$        &global update rounds before the unlearning request            \\
			$T_{unlearn}$        &global update rounds for federated unlearning,\\
			&also equals to number of stored local updates for unlearning           \\
			$\Delta t$ & interval of time between stored local updates \\
			$M_i^{local}$        & local ranking model of client $i$ before unlearning            \\
			$\Delta M_i$        & local update of client $i$ in federated learning            \\
			
			$\Delta M_{i}^{unlearn}$        & local update of client $i$ in federated unlearning            \\
			$\Delta M_{i}^{mal}$        & compromised local update of the client to be unlearned   (client $c^*$)       \\
			$z$        & parameter for poisoning attack (Section~\ref{attack})           \\
			\bottomrule
	\end{tabular}}
	\vspace{-10pt}
\end{table}

\subsection{Preliminary}
\label{pre}
Table~\ref{tbl_notation} summarises the  main notations used in this paper.
In OLTR, for a certain query $q$ and its candidate documents $\mathbf{D}$, a ranking model $M$ is used to compute a relevance score for each candidate document $d \in \mathbf{D}$. The search engine displays the documents according to their scores in descending order, and collects user's interactions with the result page. 
In a centralised setting, $M$ is iteratively trained on the server based on the features of each query-document pair and collected interaction data.

On the contrary, in FOLTR, a global ranking model $M^{global}$ is initialised in a central server and distributed to each client. At training step $t$, each client $c_i$  holds a local ranking model $M_i^{local}$ (received from the central server) and trains the local model with local data (queries, documents, interactions) for $n_i$ local updating times. After the local training phase\footnote{In our empirical study, we adapt FPDGD in which PDGD algorithm is used in the local training phase. Detailed method is specified in the original paper~\cite{wang2021effective}.}, each client sends its local update $\Delta M_{i,t} = M_i^{local} - M_t^{global}$ to a central server, that aggregates the updates from all clients to generate an updated global model. The most common aggregation rule is FedAvg~\cite{mcmahan2017communication} which updates the global model using the weighted average of all local updates:
\begin{equation}
	M_{t+1}^{global} = M_{t}^{global} + \sum_{i}^{} \frac{n_i}{\sum n_i} \Delta M_{i,t}
	\label{eq_fedagg}
\end{equation}
The newly-updated global model will be broadcast to each client and replace each local $M^{local}_i$. The whole process is repeated continuously.

\subsection{Unlearning in FOLTR}
\label{unlearn}

We next illustrate the unlearning process of FedEraser~\cite{liu2021federaser}, including how we adapted to doing unlearning for a FOLTR system.

Suppose that, during the FOLTR process and after $T$ global update rounds, a client $c^*$ requests to leave the federation and remove all the contributed local updates. Every $\Delta t$ global update rounds (i.e. at rounds $\{1,1+\Delta t,1+2\Delta t, ...\}$ etc.), we instruct each client to store their local updates $\Delta M_{i,t}$.
The number of local updates stored by each client should then be $T_{unlearn} = [\frac{T}{\Delta t}]$. 

The unlearning process takes place as outlined below, in which steps (2)-(4) are performed iteratively and the iterations correspond to the global update rounds $\{1,1+\Delta t,1+2\Delta t, ...\}$ in the original FOLTR process:

\begin{enumerate}[leftmargin=14pt]
	\item The global ranking model is initialized in the same way as in the original FOLTR process, and is passed to the clients. 
	
	\item Then, each client $c_i$ but $c^*$ (which left the federation), updates the local model by $n_i'$ steps ($n_i' < n_i$, the local step in unlearning $n_i'$ is by design smaller than that before unlearning, i.e. $n_i$). 
	
	\item Each client calibrates the local update $\Delta M_{i,t}'$ using the stored historical local update $\Delta M_{i,t}$ according to:
	\begin{equation}
		\Delta M_{i,t}^{unlearn} = ||\Delta  M_{i,t}|| \frac{\Delta M_{i,t}'}{||\Delta M_{i,t}'||}	
		\label{eq_federaser}
	\end{equation}
	where $||\Delta  M_{i,t}||$ indicates the step size of the global update and $\frac{\Delta M_{i,t}'}{||\Delta M_{i,t}'||}$ indicates the direction of the update ($|| \cdot ||$ is $L_2$-norm). 
	
	\item Each client  $c_i$ sends the calibrated updates $\Delta M_{i,t}^{unlearn}$ to the central server and the global model is updated using Equation~\ref{eq_fedagg}. 
\end{enumerate}	

After $T_{unlearn}$ times of global aggregation, the unlearned global model is obtained. In theory, the impact of client $c^*$ is still imposed through $\Delta M_{i,t}$ in Eq.~\ref{eq_federaser}. But this impact is weakened since $\Delta M_{i,t}$ is used in fewer global updating times and the way of using $\Delta M_{i,t}$ is different from the original FOLTR process (Eq.~\ref{eq_fedagg}). The actual impact will be evaluated as in Sec.~\ref{attack}.

\subsection{Efficiency Analysis}
In total, the above unlearning approach only requires $n_i' \cdot T_{unlearn}$ local updates for each client $c_i$. Retraining the federated ranker from scratch instead requires $n_i\cdot T$ updates ($n_i'$ is set to be less than $n_i$ to reduce the local computation times). Thus,  in terms of training efficiency compared to the baseline condition of retraining from scratch, the unlearning approach provides a reduction of $(n_i\cdot T)/(n_i' \cdot T_{unlearn}) = \frac{n_i}{n_i'}\cdot \Delta t$ local updates for each client $c_i$. Along with a reduction in local training, the unlearning process also reduces the communications required between  each client and the central server by $\frac{T}{T_{unlearn}} = \Delta t$ times. This reduced number of updates comes at the expense of some extra space required to store the $T_{unlearn}$ updates\footnote{These local updates would ideally be stored within each client, but they could be stored instead in the central server: this though would required extra communication cost to provide the local updates back to the clients when needed.}. For each client, the storage space cost is $[\frac{T}{\Delta t}] \cdot ||\Delta  M_{i,t}||$, where $||\Delta  M_{i,t}||$ indicates the $L_2$-norm of each local updates and it is determined by ranking model structure and the number of features.

\subsection{Evaluating Unlearning}
\label{attack}

The impact of a typical client on the whole FOLTR system can be marginal. For the purpose of evaluating the unlearning approach, we need to magnify the effect of the client that is leaving the FOLTR system and make this effect relatively controllable. In previous research on machine unlearning and federated unlearning, the effectiveness of unlearning is verified by comparing the effectiveness of the model before and after unlearning~\cite{chundawat2023zero}. This evaluation method comes with a drawback: effective unlearning is associated with a loss in model effectiveness -- a situation that is undesirable and that would penalise methods that can unlearn and do not hurt model effectiveness. 
An alternative direction has been recently proposed: leverage poisoning or backdoor attacks to evaluate unlearning~\cite{yuan2023federated,wu2022federated,halimi2022federated,sommer2020towards,sommer2022athena,hu2022membership}.

Inspired by poisoning attacks methods for federated learning~\cite{baruch2019little} and FOLTR systems~\cite{wang2023analysis} where malicious clients compromise the effectiveness of the trained models by poisoning the local training data or the model updates, we add noise to the local updates of the client to be unlearned. By doing so, we expect the unlearned clients to be clearly distinguishable from others. Due to the noise being injected, the original (i.e., before unlearning) global model performs worse to some extent compared to the model trained without the poisoned client (i.e., retrained from scratch). However, after unlearning, if the contribution of the poisoned client has been successfully removed, the overall effectiveness of the unlearned global model should improve, achieving similar effectiveness as if it was retrained from scratch. We instantiate this intuition by compromising the unlearned client's ($c^*$) local model after each local updating phase using:

\begin{equation}
M_{c^*}^{mal} = - z \cdot M_{c^*}^{local}	
\label{eq_poison_model}
\end{equation}
where $z>0$ represents how much we compromise the local model, while the negative coefficient is added to change the local model to its opposite direction. Thus, the compromised local update for the client to be unlearned is:

\begin{equation}
\Delta M_{c^*}^{mal} = - z \cdot \Delta M_{c^*}^{local} - (z+1) \cdot M_{t}^{global}	
\label{eq_poison_update}
\end{equation}
In the global updating phase before unlearning (Eq.~\ref{eq_fedagg}), we replace $\Delta M_{c^*}^{local}$ with $\Delta M_{c^*}^{mal}$ for $c^*$. In our experiments, we set $z = 2$ as we found it sufficient in degrading the effectiveness of the global model. 
Note here our poisoning method does not have the burden of hiding from detection as in real poisoning attack scenarios and thus it is simple and its parameters can be tuned as per need.

A key tenet of this evaluation is that the ability to remove such a distinguishable client is equivalent to the ability to remove a much less unique client: we are unsure whether this assumption holds true, and we are not aware of relevant literature that clearly support this.
\section{Experimental Setup}

\textbf{Datasets.} 
We evaluate using 4 common learning-to-rank (LTR) datasets: MQ2007~\cite{qin2013introducing}, MSLR-WEB10k~\cite{qin2013introducing}, Yahoo~\cite{chapelle2011yahoo}, and Istella-S~\cite{lucchese2016post}. Each dataset contains query ids and candidate document lists for each query, which is formalized as exclusive query-document pairs. Each query-document pair is represented by a multi-dimensional feature vector and annotated relevance label of the corresponding document. Among the selected datasets, MQ2007~\cite{qin2013introducing} is the smallest with 1,700 queries, 46-dimensional feature vectors, and 3-level relevance assessments (from \textit{not relevant}: 0 to \textit{very relevant}: 2). Provided by commercial search engines, the other three datasets are larger and more recent. MSLR-WEB10k has 10,000 queries and each query is associated with 125 documents on average, each represented with 136 features. Yahoo has 29,900 queries and each query-document pair has 700 features. Istella-S is the largest, with 33,018 queries, 220 features, and an average of 103 documents per query. These three commercial datasets are all annotated for relevance on a five-grade-scale: from \textit{not relevant} (0) to \textit{perfectly relevant} (4). Both MQ2007 and MSLR-WEB10k datasets have five data splits (stored separately in five data folders) while Yahoo and Istella-S contain only one. Our experimental results are averaged across all data splits.

\textbf{User simulations.} 
We follow the standard setup for user simulations in OLTR~\cite{oosterhuis2018differentiable,wang2021effective,wang2022non} by randomly selecting queries for a user and relying on the \textit{Cascade Click Model} (CCM) click model~\cite{guo2009efficient} for simulating user's clicks based the relevance label and ranking position of the candidate documents. Specifically, for each query, we limit the search engine result page (SERP) to 10 documents. User clicks on the displayed ranking list are generated based on the SDBN click model. Each user is assumed to inspect every document displayed in a SERP from top to bottom while clicks the document with click probability $P(click =1 | rel(d))$ conditioned with the actual relevance label. After a click happens, the user will stop the browsing session with stop probability $P(stop = 1 | click = 1, rel(d))$, or continue otherwise. In our experiments, same as the aforementioned previous works, we consider three widely-used instantiations of SDBN click model: \textit{perfect, navigational, informational}. A \textit{perfect} user inspects every document in a SREP with high chance to click high relevant documents thus provides very reliable feedback. The \textit{navigational} user also searches for reasonably relevant document but has a higher chance of stopping browsing after one click. The \textit{informational} 
user provides the most noisy click feedback as they do not have a specific preference on what to click and when to stop.
The implementation of these click models is detailed in Table~\ref{tbl:click}.

\textbf{Federated setup.} We consider 10 clients participating in the FOLTR process; among these 10, one client requests to be unlearned. The original federated setup (before unlearning) involves 5 local updating steps ($n_i = 5$) among all participants and 10,000 global updating steps ($T$). We assume the client requests to leave the federation at global step $T = 10,000$. During the original training in our FOLTR experiments (i.e., before the unlearning request is proposed), each client holds a copy of the current ranker and updates the local ranker by issuing $n_i = 5$ queries along with the respective click responses. After the local updating finishes, the central server will receive the updated ranker from each client and aggregate all local messages to update the global one. At each $\Delta t$ global steps (i.e. at rounds $\{1,1+\Delta t,1+2\Delta t, ...\}$ etc.), each client will also keep a copy of their local ranker update to the local device for the calibration purpose in the unlearning process. In our evaluation of unlearning (results in Figure~\ref{fig_unlearn}), we set $\Delta t = 10$ thus the total number of stored local updates is $[\frac{T}{\Delta t}] = 1000$, equaling to the global steps in the unlearning process ($T_{unlearn}$). For further hyper-parameter analysis (results in Table~\ref{tbl_unlearn}), we set a wider value ranges of $\Delta t \in \{5, 10, 20\}$. The unlearning process follows the same federated setup, with $T_{unlearn}$ global steps, $n'_i \in \{1,2,3,4\}$ local steps for the remaining 9 clients. 
We experiment with a linear model as the ranker with learning rate $\eta = 0.1$ and zero initialization, which we train with and rely on the configuration of the state-of-the-art FOLTR method, FPDGD~\cite{wang2021effective}.

\textbf{Evaluation metric.} As we limit each SERP to 10 documents, nDCG@10 is used for evaluating the overall offline effectiveness of the global ranker both before and after unlearning. Effectiveness is measured by averaging the nDCG scores of the global ranker on the queries in the held-out test dataset. This is in line with previous work on OLTR and FOLTR. Unlike previous work~\cite{oosterhuis2018differentiable,wang2021effective,jia2022learning}, online evaluation is not considered in this work, as we do not need to monitor user experience during model update. Instead, we focus on measuring the overall impact of the unlearned client (the "attacker" in Figure~\ref{fig_origin}) comparing to other baselines, and the dynamic or final performance gain during the unlearning process (results in Figure~\ref{fig_unlearn} and Table~\ref{tbl_unlearn}) following our evaluation process specifically for unlearning (specified in Section~\ref{attack}).

\newcommand{\tc}[1]{\multicolumn{1}{c}{#1}}
\setlength{\tabcolsep}{3mm}

\begin{table}[t!]
	\centering
	\caption[centre]{Instantiations of the CCM click model used to simulate user behaviour. $rel(d)$: relevance label for document $d$. In MQ2007, only three-levels of relevance are used: we report values for this in brackets.}\label{tbl:click}
	\resizebox{0.9\columnwidth}{!}{		
		\begin{tabularx}{\textwidth}{XXXXXXXXXXX}
			\toprule
			& \multicolumn{5}{c}{$P(\mathit{click}=1\mid rel(d))$} & \multicolumn{5}{c}{$P(\mathit{stop}=1\mid click=1,  rel(d))$} \\
			\cmidrule(r){2-6}  \cmidrule(){7-11}
			\emph{rel(d)} & \tc{0}& \tc{1} &\tc{2} & \tc{3}& \tc{4}&  \tc{0} & \tc{1} & \tc{2} & \tc{3} & \tc{4} \\
			\midrule
			\emph{per.}  &  0.0 (0.0)&  0.2 (0.5)&  0.4 (1.0)&  0.8 (-)&  1.0 (-)& 0.0 (0.0)& 0.0 (0.0)&  0.0 (0.0)&  0.0 (-)&  0.0 (-)\\
			\emph{nav.} & 0.05 (0.05)& 0.3 (0.5)&  0.5 (0.95)&  0.7 (-)&  0.95 (-)& 0.2 (0.2)&  0.3  (0.5)&  0.5 (0.9)&  0.7 (-)&  0.9 (-)\\
			\emph{inf.} &  0.4 (0.4)&  0.6 (0.7)&  0.7 (0.9)&  0.8 (-)&  0.9 (-)& 0.1 (0.1)&  0.2 (0.3)&  0.3 (0.5)&  0.4 (-)&  0.5 (-)\\
			\bottomrule
	\end{tabularx}}
	\vspace{-10pt}
\end{table}

\vspace{-10pt}
\section{Results and Analysis}

\subsection{Validation of Evaluation Methodology}

\begin{wrapfigure}{r}{0.5\textwidth}
	\vspace{-50pt}
	\begin{center}
		\includegraphics[width=0.5\columnwidth]{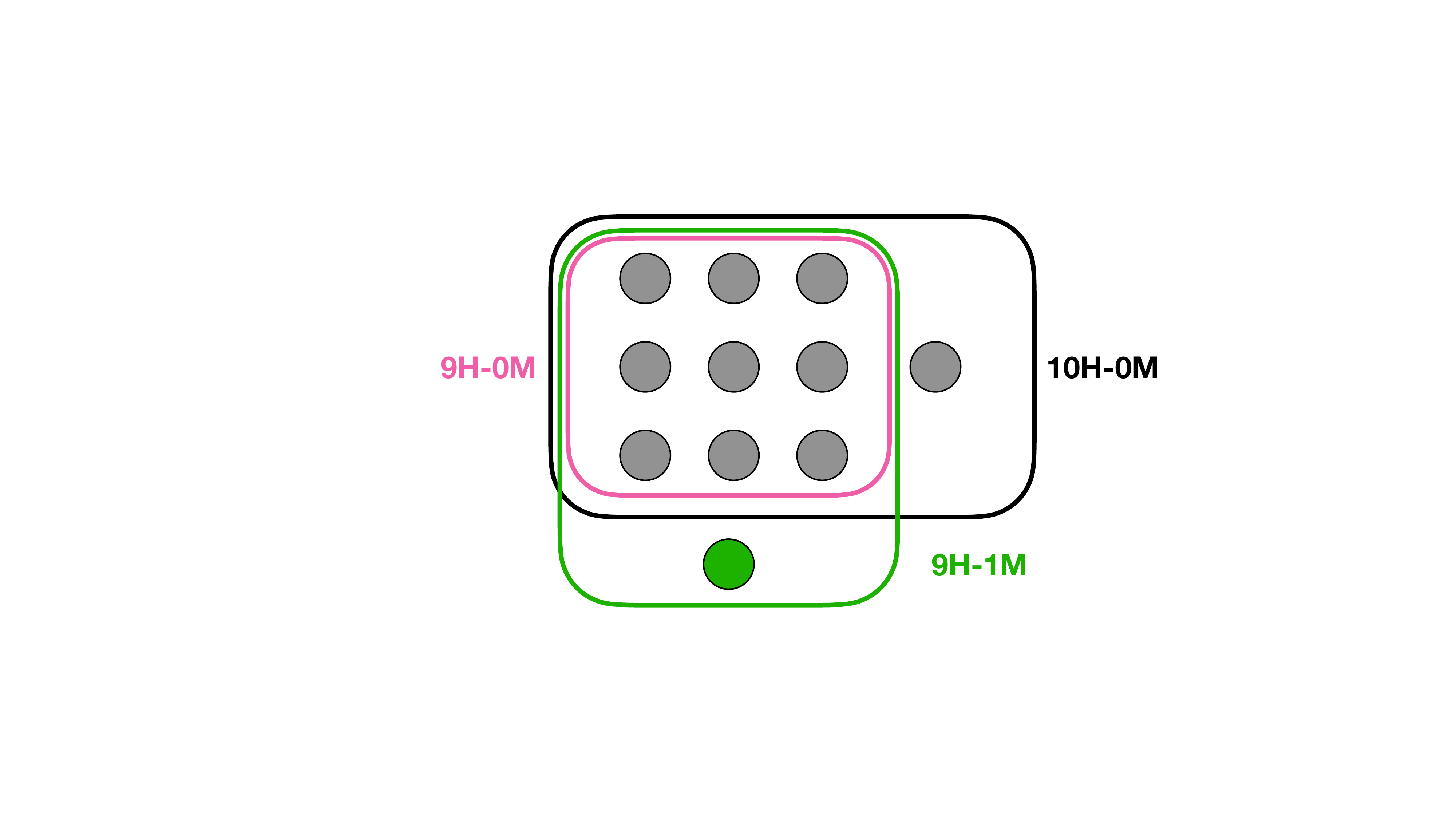}
	\end{center}
\vspace{-16pt}
	\caption{Relationships between FOLTR configurations: 9H-1M (green line), 10H-0M (black), 9H-0M (pink). Circles are clients.
		\label{fig_configs}}
	\vspace{-30pt}
\end{wrapfigure}

The unlearning evaluation methodology is based on instantiating the unlearned client as a malicious actor that injects noise in the learning process. We start by investigating if our evaluation methodology could be effective in identifying whether the unlearning has happened. For this, we consider three configurations (visualized in Figure~\ref{fig_configs}):

\begin{enumerate}[leftmargin=20pt]
	\item 9H-1M (green line): effectiveness obtained when one of the 10 clients is set to produce noisy updates (i.e. one client behaves maliciously).
	
	\item 10H-0M (black line): effectiveness obtained when all 10 clients behave in an honest way, (i.e. no client is acting maliciously).
	
	\item 9H-0M (pink line): effectiveness obtained when considering only the 9 honest clients from above, without the client that produces noisy updates (i.e. no malicious client).
\end{enumerate}

Fig.~\ref{fig_configs} clarifies the relationship between these rankers -- all rankers share the same common set of 9 honest clients, but they differ in the 10th client considered: a malicious client for 9H-1M, an honest client for 10H-0M, and no 10th client for 9H-0M.

\begin{figure*}[b]
	\centering
	%		\subfigure[\textbf{MQ2007}] {\label{fig:mq2007} \includegraphics[width=1\textwidth]{images/MQ2007_FPDGD_clients10_batch5_updates10000_FedEraser_CCM_linear_offline_ndcg.pdf}} 
	\subfigure[\textbf{MSLR-WEB10k}] {\label{fig:mslr} \includegraphics[width=1\textwidth]{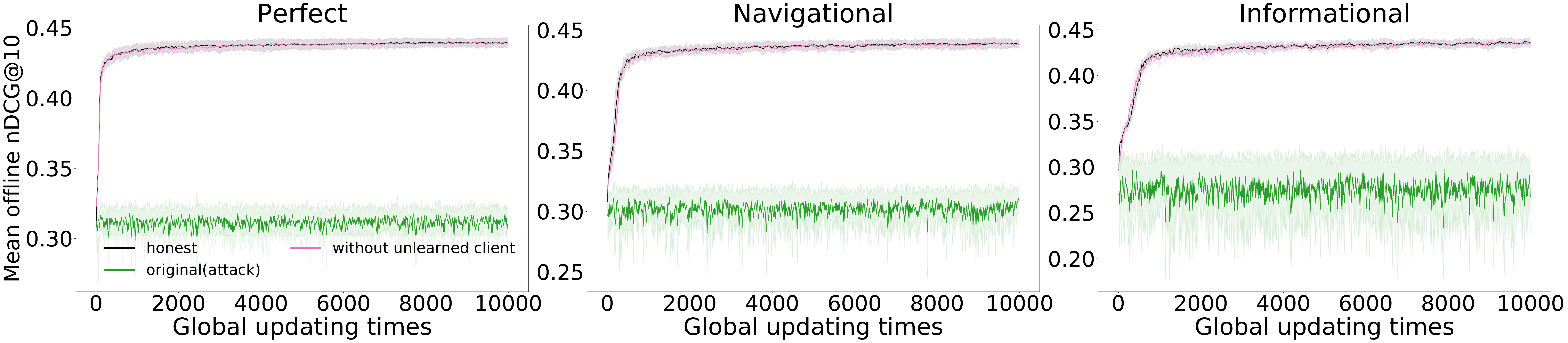}} 
	%		\subfigure[\textbf{Yahoo}] {\label{fig:yahoo} \includegraphics[width=1\textwidth]{images/Yahoo_FPDGD_clients10_batch5_updates10000_FedEraser_CCM_linear_offline_ndcg.pdf}} 
	%		\subfigure[\textbf{Istella-S}] {\label{fig:istella} \includegraphics[width=1\textwidth]{images/Istella-s_FPDGD_clients10_batch5_updates10000_FedEraser_CCM_linear_offline_ndcg.pdf}}
	%	\vspace{-12pt} 
	\caption{Offline effectiveness (nDCG@10) obtained under the 9H-1M (green line), 10H-0M (black line), 9H-0M (pink line) FOLTR configurations with three click modes (\textit{Perfect, Navigational, Informational}). Results are averaged across all dataset splits and experimental runs. These results motivate the use of the evaluation methodology based on the malicious client to evaluate the effectiveness of unlearning.
		\label{fig_origin}}
	%			\vspace{-15pt}
\end{figure*}

\begin{figure*}[t]
	\centering
	\vspace{-10pt}
	\subfigure[\textbf{MQ2007}] {\label{fig_mq2007_unlearn} \includegraphics[width=0.9\textwidth]{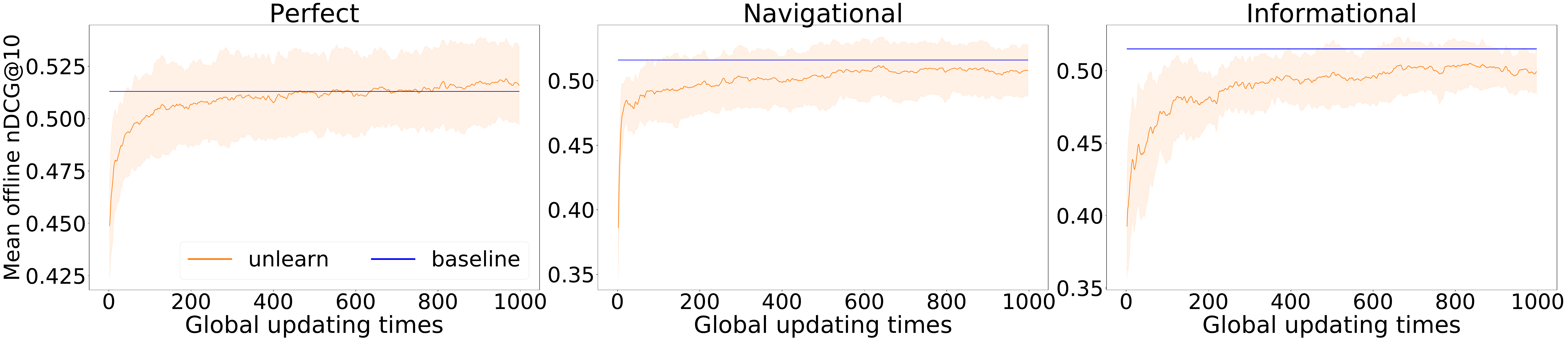}} 
	\subfigure[\textbf{MSLR-WEB10k}] {\label{fig_mslr_unlearn} \includegraphics[width=0.9\textwidth]{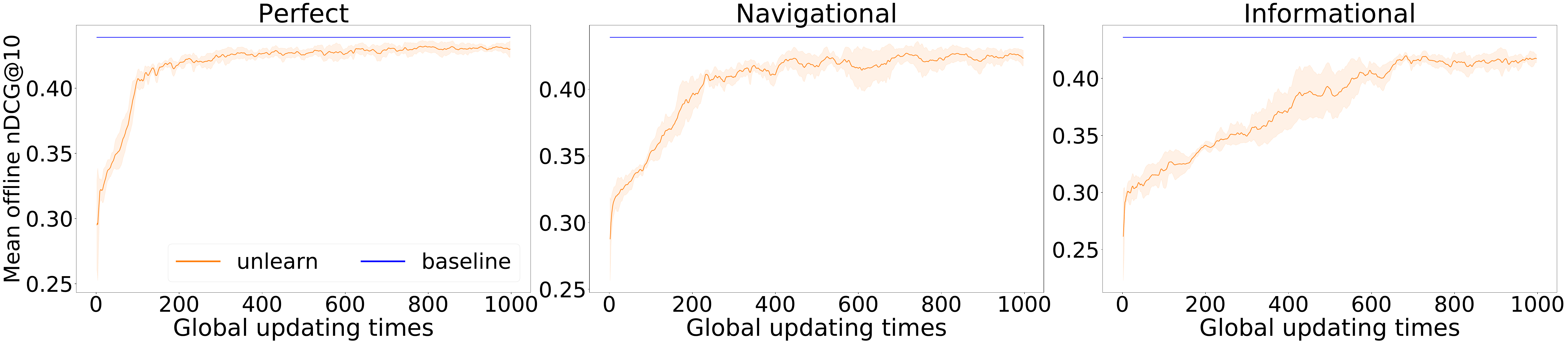}} 
	\subfigure[\textbf{Yahoo}] {\label{fig_yahoo_unlearn} \includegraphics[width=0.9\textwidth]{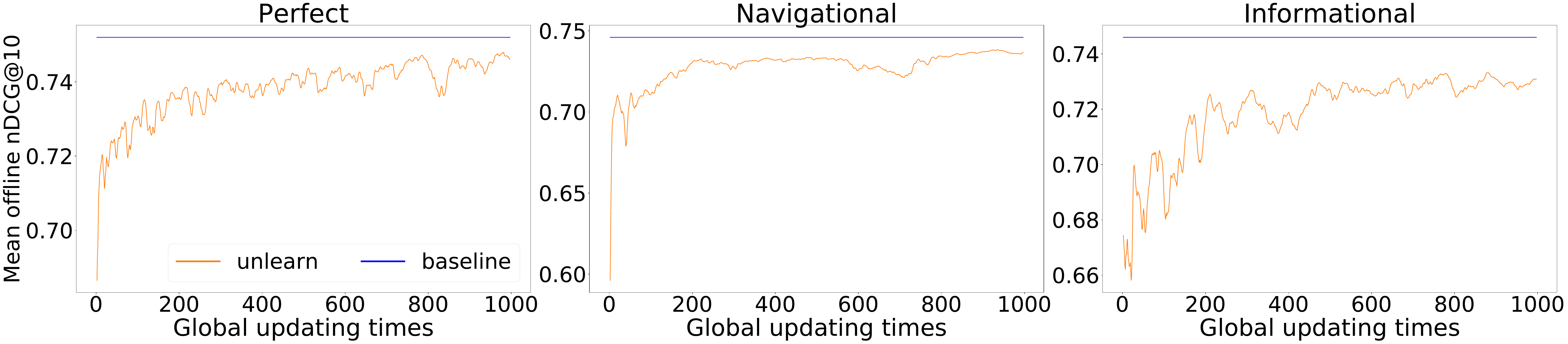}} 
	\subfigure[\textbf{Istella-S}] {\label{fig_istella_unlearn} \includegraphics[width=0.9\textwidth]{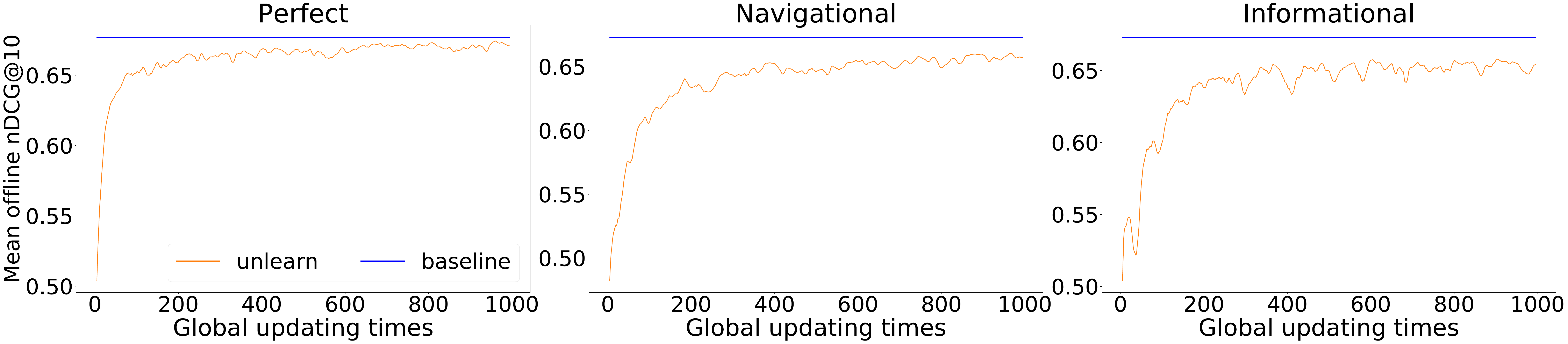}} 
	\vspace{-15pt}
	\caption{Comparison between the offline effectiveness (nDCG@10) after the unlearning method is applied (ranker $\mathcal{U}$(9H-1M) denoted as "unlearn") and the ranker is retrained from scratch after client $c^*$ is removed (ranker 9H-0M denoted as "baseline"). For the unlearning process, we set $n'_i = 3$ with $\Delta t = 10$ and show the evaluation values across all global steps. For the baseline setup, we only show the final nDCG@10 score after retraining finishes.
		\label{fig_unlearn}}
	\vspace{-15pt}
\end{figure*}

Fig.~\ref{fig_origin} reports the results obtained by these three conditions across three click modes on the MSLR-WEB10k dataset. Other datasets display similar trends; figures are omitted here for space constraints but are made available for completeness at \url{https://github.com/ielab/2024-ECIR-foltr-unlearning}. The results highlight that it is the addition of the malicious client (i.e. client $c^*$ that will be the target of the unlearning) that sensibly reduces the effectiveness of the ranker. Note that the 9H-0M is the ranker one would obtain if the unlearning process was implemented as a federated re-training of the ranker from scratch by only considering the 9 clients remaining after the removal of client $c^*$. Comparing this ranker with the 10H-0M, we further highlight the need for evaluation based on the malicious client. In fact, 10H-0M and 9H-0M only consider honest clients, but in the 9H-0M one of these clients has been removed -- but the effectiveness of two rankers is indistinguishable.

\subsection{Effectiveness of Unlearning}
\label{effectiveness_unlearn}

We now investigate the effectiveness of the unlearning method. For this, we consider ranker 9H-1M and we perform unlearning to remove client $c^*$, which is the malicious client; this leads to ranker $\mathcal{U}$(9H-1M). 
Table~\ref{tbl_unlearn} reports the effectiveness of the global model after unlearning has taken place and under different settings of hyper-parameters $n'_i$ and $\Delta t$. A more detailed analysis of the impact of $n'_i$ and $\Delta t$ is presented in Section~\ref{impact_analysis}.

In Fig.~\ref{fig_unlearn}, we report the offline effectiveness on four datasets obtained by the unlearning mechanism during the $[\frac{T}{\Delta t}] = 1,000$ global update times, where we set $\Delta t = 10$ and $n'_i = 3$.
Compared to the original model 9H-1M\footnote{The effectiveness of 9H-1M is not shown in Figure~\ref{fig_unlearn} for clarity. The reader can cross reference Figure~\ref{fig_unlearn} with Figure~\ref{fig_origin}, which instead contains the effectiveness of 9H-1M.}, the unlearned model $\mathcal{U}$(9H-1M) achieves better effectiveness and it gradually converges towards the effectiveness of the 9H-0M model, showing that the unlearned model is able to successfully remove the impact of the unlearned client $c^*$.

\subsection{Hyper-parameters Analysis}
\label{impact_analysis}

Next, we study the sensitivity of the unlearning method to its two hyper-parameters: the number of local updates for unlearning $n'_i \in \{1,2,3,4\}$ and interval of time between stored updates $\Delta t \in \{5, 10, 20\}$. We report the results of this analysis in Table~\ref{tbl_unlearn} with the final nDCG@10 score after the unlearning (or retraining). The effectiveness of ranker 9H-0H, i.e. the global model re-trained from scratch for $T = 10,000$ iterations and without client $c^*$, represents the baseline condition.

\textbf{Impact of $\mathbf{n'_i}$.} 
By design, lower values of $n'_i$ lead to higher time savings as the time required by each local update is similar.
However, lower values of $n'_i$ may mean there is insufficient training data in each iteration, and this can lead to lower ranking effectiveness. Our experimental findings display this trade-off with relative higher effectiveness obtained when more local updates ($n'_i$) are used during the unlearning process.

\textbf{Impact of $\mathbf{\Delta t}$.}
Larger values of $\Delta t$ correspond to less global updates needed for unlearning. In fact, the required global updating time is$[\frac{T}{\Delta t}]$, where in our experiments $T = 10,000$. Table~\ref{tbl_unlearn} shows that, in most cases, the model for which unlearning has taken place delivers higher ranking effectiveness for small values of $\Delta t$. This higher effectiveness, however, comes at the cost of extra time required by the unlearning process.

\begin{table*}[t]
	\centering
	\caption{Offline effectiveness (nDCG@10) of (1) the ranker trained from scratch after client $c^*$ has been removed (9H-0M), and (2) the ranker for which unlearning is performed with our method ($\mathcal{U}$(9H-1M)). Effectiveness is analysed with respect to the hyper-parameters $n'_i \in \{1,2,3,4\}$ and $\Delta t \in \{5, 10, 20\}$, under three different click models, and averaged across dataset splits. The best results are highlighted in boldface with superscripts denoting results for statistical significance study (paired Student's t-test with $p \leq 0.05$ with Bonferroni correction).
		\label{tbl_unlearn}}
	\resizebox{1\columnwidth}{!}{		
		\begin{tabular}{lcc|lll|lll|lll}
			\toprule
			&&& \multicolumn{3}{c}{Click Model: Perfect} & \multicolumn{3}{c}{Click Model: Navigational} & \multicolumn{3}{c}{Click Model: Informational}\\ 
			\midrule
			Dataset&\textbf{\#}&Ranker &$\Delta t = 5$&$\Delta t = 10$&$\Delta t = 20$ &$\Delta t = 5$&$\Delta t = 10$&$\Delta t = 20$ &$\Delta t = 5$&$\Delta t = 10$&$\Delta t = 20$  \\ 
			\midrule
			\multirow{4}{*}{MQ2007}&&9H-0M&0.513&\textbf{0.513}&\textbf{0.513}&\textbf{0.516}& \textbf{0.516}&\textbf{0.516}& \textbf{0.515}&\textbf{0.515}&\textbf{0.515}\\ 
			&a&$\mathcal{U}$(9H-1M), $n'_i=4$&0.514 &0.513&0.511&0.508&0.507&0.501&0.506&0.502&0.492\\
			&b&$\mathcal{U}$(9H-1M), $n'_i=3$& 0.514&0.513&0.507&0.510&0.504&0.499&0.505&0.498&0.491\\
			&c&$\mathcal{U}$(9H-1M), $n'_i=2$&\textbf{0.515} &0.510&0.507&0.505&0.499&0.499&0.502&0.497&0.483\\
			&d&$\mathcal{U}$(9H-1M), $n'_i=1$& 0.511&0.506&0.501&0.504&0.498&0.496&0.489&0.491&0.484\\
			
			\midrule
			\multirow{4}{*}{MSLR-10k}&&9H-0M&\textbf{0.439}&\textbf{0.439}$^{d}$&\textbf{0.439}$^{cd}$&\textbf{0.439}$^{d}$&\textbf{0.439}$^{cd}$&\textbf{0.439}$^{abcd}$&\textbf{0.436}$^{cd}$&\textbf{0.436}$^{abcd}$&\textbf{0.436}$^{abcd}$\\ 
			&a&$\mathcal{U}$(9H-1M), $n'_i=4$&0.436&0.433&0.429&0.433&0.428&0.422&0.429&0.413&0.389\\
			&b&$\mathcal{U}$(9H-1M), $n'_i=3$&0.435 &0.432&0.427&0.429& 0.428&0.414&0.424&0.412&0.361\\
			&c&$\mathcal{U}$(9H-1M), $n'_i=2$&0.432 &0.431&0.423&0.426&0.421&0.389&0.420&0.386&0.348\\ 
			&d&$\mathcal{U}$(9H-1M), $n'_i=1$&0.430 &0.422 &0.414&0.418&0.382&0.344&0.393&0.343&0.332\\
			
			\midrule
			\multirow{4}{*}{Yahoo}&&9H-0M& \textbf{0.752}&\textbf{0.752}$^{bcd}$&\textbf{0.752}$^{abcd}$&\textbf{0.746}$^{bd}$&\textbf{0.746}$^{bcd}$&\textbf{0.746}$^{abcd}$&\textbf{0.746}$^{abcd}$&\textbf{0.746}$^{abcd}$&\textbf{0.746}$^{abcd}$\\
			&a&$\mathcal{U}$(9H-1M), $n'_i=4$&0.748 &0.746&0.741&0.743&0.743&0.734&0.738&0.725&0.726\\ 
			&b&$\mathcal{U}$(9H-1M), $n'_i=3$&0.748&0.740&0.742&0.738&0.739&0.735&0.736&0.725&0.721\\ 
			&c&$\mathcal{U}$(9H-1M), $n'_i=2$&0.746 &0.746&0.742&0.738&0.735&0.733&0.732&0.725&0.715\\
			&d&$\mathcal{U}$(9H-1M), $n'_i=1$&0.746 &0.739&0.732&0.734&0.726&0.725&0.730&0.720&0.713\\
			
			\midrule
			\multirow{4}{*}{Istella-s}&&9H-0M& \textbf{0.677}$^{cd}$&\textbf{0.677}$^{cd}$&\textbf{0.677}$^{abcd}$&\textbf{0.673}$^{bcd}$ &\textbf{0.673}$^{abcd}$ &\textbf{0.673}$^{abcd}$ &\textbf{0.673}$^{abcd}$&\textbf{0.673}$^{abcd}$&\textbf{0.673}$^{abcd}$\\
			&a&$\mathcal{U}$(9H-1M), $n'_i=4$&0.673 & 0.672 &0.664&0.665& 0.659 &0.650&0.663& 0.655&0.654\\
			&b&$\mathcal{U}$(9H-1M), $n'_i=3$&0.673 & 0.670 &0.663&0.664& 0.662 &0.643&0.664& 0.658 &0.639\\
			&c&$\mathcal{U}$(9H-1M), $n'_i=2$&0.669 & 0.670 &0.663&0.661& 0.652 &0.637&0.659& 0.653&0.635\\
			&d&$\mathcal{U}$(9H-1M), $n'_i=1$&0.669 &0.662 &0.657&0.653& 0.637 &0.604&0.649& 0.641&0.600\\
			
			\bottomrule
	\end{tabular}}
\end{table*}

\vspace{-9pt}
\section{Conclusion}

This paper is the first study that investigates unlearning in Federated Online Learning to Rank. For this, we adapt the FedEraser method~\cite{liu2021federaser}, developed for general federated learning problems, to the unique context of federated online learning to rank, where rankers are learned in an online manner on implicit interactions (e.g. clicks). 
We further modify the method to leverage stored historic local updates to guide and accelerate the process of unlearning. To evaluate the effectiveness of the unlearning method, we adapt the idea of poisoning attacks to the context of determining whether the contributions of a client to be unlearned made to the ranker are effectively erased from the ranker itself. Experimental results on four popular LTR datasets show both the effectiveness and the efficiency of the unlearning method. In particular: (1) the search effectiveness of the global model once unlearning takes place converges to the effectiveness of the ranker if the removed client did not take part in the FOLTR system from the start, (2) the contributions made to the ranker by the unlearned client are effectively removed, and (3) less local and global steps are required by unlearning compared to retraining the model from scratch. 

%
%\subsubsection{Acknowledgements} Please place your acknowledgments at
%the end of the paper, preceded by an unnumbered run-in heading (i.e.
%3rd-level heading).

\bibliographystyle{splncs04}
\bibliography{ecir-foltr-unlearning}

\begin{thebibliography}{10}
\providecommand{\url}[1]{\texttt{#1}}
\providecommand{\urlprefix}{URL }
\providecommand{\doi}[1]{https://doi.org/#1}

\bibitem{ai2021unbiased}
Ai, Q., Yang, T., Wang, H., Mao, J.: Unbiased learning to rank: online or
  offline? ACM Transactions on Information Systems (TOIS)  \textbf{39}(2),
  1--29 (2021)

\bibitem{DBLP:journals/access/AldaghriMB21}
Aldaghri, N., Mahdavifar, H., Beirami, A.: Coded machine unlearning. {IEEE}
  Access  \textbf{9},  88137--88150 (2021). \doi{10.1109/ACCESS.2021.3090019},
  \url{https://doi.org/10.1109/ACCESS.2021.3090019}

\bibitem{baruch2019little}
Baruch, G., Baruch, M., Goldberg, Y.: A little is enough: Circumventing
  defenses for distributed learning. Advances in Neural Information Processing
  Systems  \textbf{32} (2019)

\bibitem{DBLP:journals/ml/BaumhauerSZ22}
Baumhauer, T., Sch{\"{o}}ttle, P., Zeppelzauer, M.: Machine unlearning: linear
  filtration for logit-based classifiers. Mach. Learn.  \textbf{111}(9),
  3203--3226 (2022). \doi{10.1007/s10994-022-06178-9},
  \url{https://doi.org/10.1007/s10994-022-06178-9}

\bibitem{bourtoule2021machine}
Bourtoule, L., Chandrasekaran, V., Choquette-Choo, C.A., Jia, H., Travers, A.,
  Zhang, B., Lie, D., Papernot, N.: Machine unlearning. In: 2021 IEEE Symposium
  on Security and Privacy (SP). pp. 141--159. IEEE (2021)

\bibitem{DBLP:conf/icml/BrophyL21}
Brophy, J., Lowd, D.: Machine unlearning for random forests. In: Meila, M.,
  Zhang, T. (eds.) Proceedings of the 38th International Conference on Machine
  Learning, {ICML} 2021, 18-24 July 2021, Virtual Event. Proceedings of Machine
  Learning Research, vol.~139, pp. 1092--1104. {PMLR} (2021),
  \url{http://proceedings.mlr.press/v139/brophy21a.html}

\bibitem{DBLP:conf/sp/CaoY15}
Cao, Y., Yang, J.: Towards making systems forget with machine unlearning. In:
  2015 {IEEE} Symposium on Security and Privacy, {SP} 2015, San Jose, CA, USA,
  May 17-21, 2015. pp. 463--480. {IEEE} Computer Society (2015).
  \doi{10.1109/SP.2015.35}, \url{https://doi.org/10.1109/SP.2015.35}

\bibitem{chapelle2011yahoo}
Chapelle, O., Chang, Y.: Yahoo! learning to rank challenge overview. In:
  Proceedings of the learning to rank challenge. pp. 1--24. PMLR (2011)

\bibitem{che2023fast}
Che, T., Zhou, Y., Zhang, Z., Lyu, L., Liu, J., Yan, D., Dou, D., Huan, J.:
  Fast federated machine unlearning with nonlinear functional theory. In:
  International conference on machine learning. pp. 4241--4268. PMLR (2023)

\bibitem{chen2021machine}
Chen, M., Zhang, Z., Wang, T., Backes, M., Humbert, M., Zhang, Y.: When machine
  unlearning jeopardizes privacy. In: Proceedings of the 2021 ACM SIGSAC
  conference on computer and communications security. pp. 896--911 (2021)

\bibitem{DBLP:conf/ccs/Chen000H022}
Chen, M., Zhang, Z., Wang, T., Backes, M., Humbert, M., Zhang, Y.: Graph
  unlearning. In: Yin, H., Stavrou, A., Cremers, C., Shi, E. (eds.) Proceedings
  of the 2022 {ACM} {SIGSAC} Conference on Computer and Communications
  Security, {CCS} 2022, Los Angeles, CA, USA, November 7-11, 2022. pp.
  499--513. {ACM} (2022). \doi{10.1145/3548606.3559352},
  \url{https://doi.org/10.1145/3548606.3559352}

\bibitem{DBLP:journals/cluster/ChenXXZ19}
Chen, Y., Xiong, J., Xu, W., Zuo, J.: A novel online incremental and
  decremental learning algorithm based on variable support vector machine.
  Clust. Comput.  \textbf{22}(Supplement),  7435--7445 (2019).
  \doi{10.1007/s10586-018-1772-4},
  \url{https://doi.org/10.1007/s10586-018-1772-4}

\bibitem{chundawat2023zero}
Chundawat, V.S., Tarun, A.K., Mandal, M., Kankanhalli, M.: Zero-shot machine
  unlearning. IEEE Transactions on Information Forensics and Security  (2023)

\bibitem{gao2022verifi}
Gao, X., Ma, X., Wang, J., Sun, Y., Li, B., Ji, S., Cheng, P., Chen, J.:
  Verifi: Towards verifiable federated unlearning. arXiv preprint
  arXiv:2205.12709  (2022)

\bibitem{DBLP:conf/nips/GinartGVZ19}
Ginart, A., Guan, M.Y., Valiant, G., Zou, J.: Making {AI} forget you: Data
  deletion in machine learning. In: Wallach, H.M., Larochelle, H., Beygelzimer,
  A., d'Alch{\'{e}}{-}Buc, F., Fox, E.B., Garnett, R. (eds.) Advances in Neural
  Information Processing Systems 32: Annual Conference on Neural Information
  Processing Systems 2019, NeurIPS 2019, December 8-14, 2019, Vancouver, BC,
  Canada. pp. 3513--3526 (2019),
  \url{https://proceedings.neurips.cc/paper/2019/hash/cb79f8fa58b91d3af6c9c991f63962d3-Abstract.html}

\bibitem{guo2009efficient}
Guo, F., Liu, C., Wang, Y.M.: Efficient multiple-click models in web search.
  In: Proceedings of the second acm international conference on web search and
  data mining. pp. 124--131 (2009)

\bibitem{halimi2022federated}
Halimi, A., Kadhe, S., Rawat, A., Baracaldo, N.: Federated unlearning: How to
  efficiently erase a client in fl? arXiv preprint arXiv:2207.05521  (2022)

\bibitem{hofmann2013reusing}
Hofmann, K., Schuth, A., Whiteson, S., De~Rijke, M.: Reusing historical
  interaction data for faster online learning to rank for ir. In: Proceedings
  of the sixth ACM international conference on Web search and data mining. pp.
  183--192 (2013)

\bibitem{hu2022membership}
Hu, H., Salcic, Z., Dobbie, G., Chen, J., Sun, L., Zhang, X.: Membership
  inference via backdooring. In: The 31st International Joint Conference on
  Artificial Intelligence (IJCAI-22) (2022)

\bibitem{jia2022learning}
Jia, Y., Wang, H.: Learning neural ranking models online from implicit user
  feedback. In: Proceedings of the ACM Web Conference 2022. pp. 431--441 (2022)

\bibitem{kairouz2021advances}
Kairouz, P., McMahan, H.B., Avent, B., Bellet, A., Bennis, M., Bhagoji, A.N.,
  Bonawitz, K., Charles, Z., Cormode, G., Cummings, R., et~al.: Advances and
  open problems in federated learning. Foundations and Trends{\textregistered}
  in Machine Learning  \textbf{14}(1--2),  1--210 (2021)

\bibitem{kharitonov2019federated}
Kharitonov, E.: Federated online learning to rank with evolution strategies.
  In: Proceedings of the Twelfth ACM International Conference on Web Search and
  Data Mining. pp. 249--257 (2019)

\bibitem{liu2021federaser}
Liu, G., Ma, X., Yang, Y., Wang, C., Liu, J.: Federaser: Enabling efficient
  client-level data removal from federated learning models. In: 2021 IEEE/ACM
  29th International Symposium on Quality of Service (IWQOS). pp. 1--10. IEEE
  (2021)

\bibitem{DBLP:journals/corr/abs-2003-10933}
Liu, Y., Ma, Z., Liu, X., Ma, J.: Learn to forget: User-level memorization
  elimination in federated learning. CoRR  \textbf{abs/2003.10933} (2020),
  \url{https://arxiv.org/abs/2003.10933}

\bibitem{DBLP:conf/infocom/LiuXYWL22}
Liu, Y., Xu, L., Yuan, X., Wang, C., Li, B.: The right to be forgotten in
  federated learning: An efficient realization with rapid retraining. In:
  {IEEE} {INFOCOM} 2022 - {IEEE} Conference on Computer Communications, London,
  United Kingdom, May 2-5, 2022. pp. 1749--1758. {IEEE} (2022).
  \doi{10.1109/INFOCOM48880.2022.9796721},
  \url{https://doi.org/10.1109/INFOCOM48880.2022.9796721}

\bibitem{lucchese2016post}
Lucchese, C., Nardini, F.M., Orlando, S., Perego, R., Silvestri, F., Trani, S.:
  Post-learning optimization of tree ensembles for efficient ranking. In:
  Proceedings of the 39th International ACM SIGIR conference on Research and
  Development in Information Retrieval. pp. 949--952 (2016)

\bibitem{de2020european}
de~Magalh{\~a}es, S.T.: The european union's general data protection regulation
  (gdpr). In: CYBER SECURITY PRACTITIONER'S GUIDE, pp. 529--558. World
  Scientific (2020)

\bibitem{mahadevan2021certifiable}
Mahadevan, A., Mathioudakis, M.: Certifiable machine unlearning for linear
  models. arXiv preprint arXiv:2106.15093  (2021)

\bibitem{mcmahan2017communication}
McMahan, B., Moore, E., Ramage, D., Hampson, S., y~Arcas, B.A.:
  Communication-efficient learning of deep networks from decentralized data.
  In: Artificial intelligence and statistics. pp. 1273--1282. PMLR (2017)

\bibitem{nguyen2022survey}
Nguyen, T.T., Huynh, T.T., Nguyen, P.L., Liew, A.W.C., Yin, H., Nguyen, Q.V.H.:
  A survey of machine unlearning. arXiv preprint arXiv:2209.02299  (2022)

\bibitem{oosterhuis2018differentiable}
Oosterhuis, H., de~Rijke, M.: Differentiable unbiased online learning to rank.
  In: Proceedings of the 27th ACM international conference on information and
  knowledge management. pp. 1293--1302 (2018)

\bibitem{qin2013introducing}
Qin, T., Liu, T.Y.: Introducing letor 4.0 datasets. arXiv preprint
  arXiv:1306.2597  (2013)

\bibitem{salimans2017evolution}
Salimans, T., Ho, J., Chen, X., Sidor, S., Sutskever, I.: Evolution strategies
  as a scalable alternative to reinforcement learning. arXiv preprint
  arXiv:1703.03864  (2017)

\bibitem{schuth2016multileave}
Schuth, A., Oosterhuis, H., Whiteson, S., de~Rijke, M.: Multileave gradient
  descent for fast online learning to rank. In: proceedings of the ninth ACM
  international conference on web search and data mining. pp. 457--466 (2016)

\bibitem{shokri2017membership}
Shokri, R., Stronati, M., Song, C., Shmatikov, V.: Membership inference attacks
  against machine learning models. In: 2017 IEEE symposium on security and
  privacy (SP). pp. 3--18. IEEE (2017)

\bibitem{sommer2020towards}
Sommer, D.M., Song, L., Wagh, S., Mittal, P.: Towards probabilistic
  verification of machine unlearning. arXiv preprint arXiv:2003.04247  (2020)

\bibitem{sommer2022athena}
Sommer, D.M., Song, L., Wagh, S., Mittal, P.: Athena: Probabilistic
  verification of machine unlearning. Proceedings on Privacy Enhancing
  Technologies  \textbf{3},  268--290 (2022)

\bibitem{wang2019variance}
Wang, H., Kim, S., McCord-Snook, E., Wu, Q., Wang, H.: Variance reduction in
  gradient exploration for online learning to rank. In: Proceedings of the 42nd
  International ACM SIGIR Conference on Research and Development in Information
  Retrieval. pp. 835--844 (2019)

\bibitem{wang2018efficient}
Wang, H., Langley, R., Kim, S., McCord-Snook, E., Wang, H.: Efficient
  exploration of gradient space for online learning to rank. In: The 41st
  international ACM SIGIR conference on research \& development in information
  retrieval. pp. 145--154 (2018)

\bibitem{wang2022federated}
Wang, J., Guo, S., Xie, X., Qi, H.: Federated unlearning via
  class-discriminative pruning. In: Proceedings of the ACM Web Conference 2022.
  pp. 622--632 (2022)

\bibitem{wang2021effective}
Wang, S., Liu, B., Zhuang, S., Zuccon, G.: Effective and privacy-preserving
  federated online learning to rank. In: Proceedings of the 2021 ACM SIGIR
  international conference on theory of information retrieval. pp. 3--12 (2021)

\bibitem{wang2021federated}
Wang, S., Zhuang, S., Zuccon, G.: Federated online learning to rank with
  evolution strategies: A reproducibility study. In: European Conference on
  Information Retrieval (2021)

\bibitem{wang2022non}
Wang, S., Zuccon, G.: Is non-iid data a threat in federated online learning to
  rank? In: Proceedings of the 45th International ACM SIGIR Conference on
  Research and Development in Information Retrieval. pp. 2801--2813 (2022)

\bibitem{wang2023analysis}
Wang, S., Zuccon, G.: An analysis of untargeted poisoning attack and defense
  methods for federated online learning to rank systems. In: Proceedings of the
  2023 ACM SIGIR International Conference on Theory of Information Retrieval.
  pp. 215--224 (2023)

\bibitem{wu2022federated}
Wu, C., Zhu, S., Mitra, P.: Federated unlearning with knowledge distillation.
  arXiv preprint arXiv:2201.09441  (2022)

\bibitem{yuan2023federated}
Yuan, W., Yin, H., Wu, F., Zhang, S., He, T., Wang, H.: Federated unlearning
  for on-device recommendation. In: Proceedings of the Sixteenth ACM
  International Conference on Web Search and Data Mining. pp. 393--401 (2023)

\bibitem{yue2009interactively}
Yue, Y., Joachims, T.: Interactively optimizing information retrieval systems
  as a dueling bandits problem. In: Proceedings of the 26th Annual
  International Conference on Machine Learning. pp. 1201--1208 (2009)

\bibitem{zhao2016constructing}
Zhao, T., King, I.: Constructing reliable gradient exploration for online
  learning to rank. In: Proceedings of the 25th ACM International on Conference
  on Information and Knowledge Management. pp. 1643--1652 (2016)

\bibitem{zhuang2022reinforcement}
Zhuang, S., Qiao, Z., Zuccon, G.: Reinforcement online learning to rank with
  unbiased reward shaping. Information Retrieval Journal  \textbf{25}(4),
  386--413 (2022)

\bibitem{zhuang2020counterfactual}
Zhuang, S., Zuccon, G.: Counterfactual online learning to rank. In: Advances in
  Information Retrieval: 42nd European Conference on IR Research, ECIR 2020,
  Lisbon, Portugal, April 14--17, 2020, Proceedings, Part I 42. pp. 415--430.
  Springer (2020)

\end{thebibliography}

\end{document}